\documentclass[11pt]{article}
\usepackage[sc]{mathpazo} %Like Palatino with extensive math support
\usepackage{fullpage}
\usepackage[numbers,square,sort&compress]{natbib}
\usepackage[utf8]{inputenc}
\usepackage{lineno}
\usepackage{titlesec}
\titleformat{\section}[block]{\Large\bfseries\filcenter}{\thesection}{1em}{}
\titleformat{\subsection}[block]{\Large\itshape\filcenter}{\thesubsection}{1em}{}
\titleformat{\subsubsection}[block]{\large\itshape\bfseries}{\thesubsubsection}{1em}{}
\titleformat{\paragraph}[runin]{\itshape}{\theparagraph}{1em}{}[. ]
\usepackage{graphicx}
\usepackage{amsmath,amssymb}
\usepackage{breakcites}
\usepackage{float}
\usepackage{hyperref}
\usepackage[usenames, dvipsnames]{color}

\title{Evolution of female choice under intralocus sexual conflict and genotype-by-environment interactions}

\author{Xiang-Yi Li$^{1,\ast}$ \& Luke Holman$^{2}$}
\date{}

\begin{document}

\maketitle

\noindent{} 1. Department of Evolutionary Biology and Environmental Studies, University of Zurich, Winterthurerstrasse 190, CH-8057 Zurich, Switzerland.

\noindent{} 2. School of BioSciences, University of Melbourne, Parkville, VIC 3010, Australia.

\noindent{} $\ast$ Corresponding author; E-mail: xiangyi.li@ieu.uzh.ch.
%\bigskip
%
%\textit{Manuscript elements}: Figure~1, figure~2, table~1, online appendices~A and B (including figure~A1 and figure~A2). Figure~2 is to print in color.

\bigskip\bigskip\bigskip\bigskip

\noindent{}\textbf{Keywords}: Environmental stochasticity, GEI and GxE, Lek paradox, Local adaptation, Mate preference, Sexual antagonism.

\bigskip\bigskip

\noindent{}\textbf{Manuscript type}: Research. 

\bigskip\bigskip\bigskip\bigskip\bigskip\bigskip\bigskip

\noindent{}\textbf{Data Availability}: The Matlab/Octave Simulation codes and the Mathematica Notebooks for producing the figures are available in \href{https://github.com/XiangyiLi/Evolution-of-female-choice-under-intralocus-sexual-conflict-and-genotype-by-environment-interactions.git}{\textcolor{blue}{this GitHub repository}}.

\bigskip\bigskip\bigskip\bigskip\bigskip\bigskip\bigskip\bigskip

\noindent{\footnotesize Manuscript accepted in \textit{Philosophical Transactions of the Royal Society B}: doi:10.1098/rstb.2017.0425}

%\linenumbers{}
%\modulolinenumbers[3]

\newpage{}

\section*{Abstract}

In many species, females are hypothesised to obtain ‘good genes’ for their offspring by mating with males in good condition. 
However, female preferences might deplete genetic variance and make choice redundant. 
Additionally, high-condition males sometime produce low-fitness offspring, for example because of environmental turnover and gene-by-environment interactions (GEIs) for fitness, 
or because fit males carry sexually-antagonistic alleles causing them to produce unfit daughters. 
Here, we extend previous theory by investigating the evolution of female mate choice in a spatially explicit evolutionary simulation implementing both GEIs and intralocus sexual conflict (IASC), 
under sex-specific hard or soft selection.
We show that IASC can weaken female preferences for high-condition males or even cause a preference for males in low condition, 
depending on the relative benefits of producing well-adapted sons versus daughters, which in turn depends on the relative hardness of selection on males and females. 
We discuss the relevance of our results to conservation genetics and empirical evolutionary biology.

\newpage{}

\section*{Introduction}
Selection resulting from mate choice preferences can profoundly affect evolution and demography, and can favour striking adaptations such as song, dance and elaborated coloration \citep{andersson:book:1994}.
One hypothesis is that mate choice preferences evolved to allow the choosy sex (hereafter 'females' for brevity) to select mates carrying 'good genes' (using associated phenotypic cues, such as condition-dependent sexual signals \citep{vonSchantz:PRSB:1999}), such that alleles encoding mate choice will be found in high-fitness females more often than alleles encoding random mating \citep{johnstone:BR:1995}.

A perennial problem for good genes models is that mate preferences tend to deplete genetic variance, making choice redundant and selecting against choosiness whenever it has a cost (the 'lek paradox', e.g. \citep{rowe:PRSB:1996}).
Another problem is that the notion of good genes is sugjective.
Because gene-by-environment interactions (GEIs) are ubiquitous \citep{hunt:book:2014}, many alleles only confer high fitness in certain environments.
Choosy females may therefore obtain alleles that were advantageous in the male's environment, but which are disadvantageous in the environment experienced by their offspring, due to spatial and/or temporal environmental turnover.
Similarly, many alleles confer high fitness when expressed in the 'environment' of a male body, but low fitness in a female body, or vice versa \citep{bonduriansky:TREE:2009,pennell:EE:2013}.
Such intralocus sexual conflict (IASC) means that females selecting high-condition males may therefore produce fit sons but unfit daughters, reducing or reversing the net fitness of mate choice \citep{pischedda:PLOSBIO:2006}.

Alternatively, IASC might sometimes drive the evolution of female choice. Seger \& Trivers (1986) \citep{seger:Nature:1986} showed that females can evolve to prefer males carrying female-beneficial/male-detrimental alleles, due to either a build-up of linkage disequilibrium between female preference and sexual antagonistic loci, or partial population mixing by migration between demes.
Albert \& Otto (2005) \citep{albert:Science:2005} studied the effects of IASC on female choice evolution under different sex-determination systems.
They found that if both the IASC locus and the female choice locus are autosomal, female choice cannot evolve.
% But if the IASC locus is X-linked, females should evolve to prefer low-condition males. 
%Under Z-linked IASC, females evolved to prefer low-condition or high-condition males, depending on whether the female choice locus is autosomal or Z-linked, respectively.

Similarly, modelling work has shown that GEIs can both help and hinder the spread of alleles encoding female preference \citep{kokko:genetica:2008, holman:chapter:2014}.
However, few if any previous models have simultaneously examined the effects of IASC and GEIs on female choice evolution.
Since GEIs help maintain genetic variance, they help resolve the lek paradox by maintaining locally maladapted alleles.
Conversely, GEIs lead to 'mistakes' whereby females choose a male whose alleles are poorly adapted to the environment that their offspring will experience, reducing the benefits of being choosy.
Interestingly these two effects do not cancel out: occasional mistakes actually promote the evolution of mate choice, since they help maintain a pool of maladapted males for choosy females to avoid.

We hypothesise that when individual condition depends on both local adaptation and IASC, a female preference for high-condition males may yield locally-adapted offspring, but runs the risk of producing unfit daughters. 
An additional complication is that the proportion of genetic variance in fitness that is sexually antagonistic is not fixed, but varies dynamically in response to processes such as migration, local adaptation, environmental changes, and assortative mating (e.g. empirical evidence: \citep{long:CB:2012, collet:Evolution:2016, holman:JEB:2017, martinossi:evolution:2018}; theory: \citep{arnqvist:evolution:2011, connallon:evolution:2016}).

Here, we use individual-based simulations to jointly study the evolution of female mate preferences, local adaptation under GEIs, IASC, and sex-specific (potentially co-evolving) dispersal behaviour.
Our principal aims are to test whether female preferences evolve more or less easily in populations carrying sexually antagonistic alleles, and to determine whether the conclusions of past work regarding mate choice under GEIs are robust to the addition of coevolving sexual conflict loci.
We consider a range of assumptions about sex-specific dispersal, the scale of competition, and the relative importance of IASC and local adaptation to individual condition.

\section*{Methods}
\subsubsection*{The meta-population}
We model a population of sexually-reproducing, dioecious haploids subdivided over $K$ discrete habitat patches.
The habitat patches are arranged linearly in a ring, such that every patch has two neighbours and there are no 'edge effects'. 
Each patch has an environmental state $E$, which can be thought of as an ecological variable that affects fitness.
Each individual has a 'condition' (range: 0-1), which affects the fecundity of females and the mating probability of males (provided at least one choosy female is present).
Each individual carries a 'local adaptation' locus, which determines the value of $E$ that maximises that individual's condition. 
Each individual also carries a 'sexual conflict' locus (which affects condition in a sex-specific manner), two loci controlling sex-specific dispersal, and two loci controlling female preferences, for a total of six loci.

\subsubsection*{The environment}
We create four different types of environment (see Electronic Supplementary Information [ESM] section A for detailed methods and sample time series), where 1) $E$ values are static and homogeneous across habitat patches, 2) $E$ values are static but heterogeneous across habitats, 3) $E$ values are heterogeneous and fluctuate mildly across patches, or 4) $E$ values are heterogeneous and fluctuate strongly across patches. 
In the last two types, $E$ values are temporally and spatially autocorrelated such that nearby patches and time points tend to have similar $E$.

\subsubsection*{Condition, mate choice and reproduction}
We model discrete generations in which individuals are born, disperse, reproduce and then die.
We assume that an individual's condition is determined by 1) the match between its genotype at the local adaptation locus and the environment(s) it encounters, and 2) its genotype at the sexual conflict locus. 
Specifically, the condition ($\xi$) of an individual in the reproduction phase is a weighted average of its conditions in the natal and breeding patches.
In Eqn.[\ref{eqn:condition}], the subscripts $N$ and $B$ represent the effects on condition of the environmental states in the $natal$ and $breeding$ patches respectively, 
while $C$ ('concordant') and $A$ ('antagonistic') represent the effects on condition of the local adaptation and sexual conflict loci.
The relative importance of the natal \textit{vs.} breeding habitats in determining condition is weighted by $k_{\text{natal}}$, while $k_{LA}$ ('local adaptation') controls the realtive contributions of local adaptation \textit{vs.} sexual conflict to individual condition. 
The condition of an individual at the time of breeding is then 

\begin{equation}
\xi=k_{\text{natal}}(k_{LA}\xi_{N\cdot C}+(1-k_{LA})\xi_{N\cdot A})+(1-k_{\text{natal}})(k_{LA}\xi_{B\cdot C}+(1-k_{LA})\xi_{B\cdot A}).
\label{eqn:condition}
\end{equation}

\bigskip
For the local adaptation locus, condition is maximised when the individual's genotype ($\phi$) exactly matches the environmental state $E$, and decreases exponentially as the phenotype deviates from it, so that $\xi_{i\cdot C}=\text{Exp}(-10|\phi-E|)$, where $i$ can take either $N$ (on natal patch) or $B$ (on breeding patch).
For the sexual conflict locus, we consider two different causes of sexual antagonistic selection. 
In the first scenario, sexual conflict arises from IASC, and we assume that an individual with allelic value $x$ at the locus has the condition component $\xi_{i\cdot A}=x$ if the individual is male, and $\xi_{i\cdot A}=1-x \ (i \in \{N,B\})$ if the individual is female.
In the second scenario, sexual conflict arises from different environmental optima for males and females. 
In this case, we consider an independent environmental state $E'$ (details in the ESM section A), and males have the condition component $\xi_{i\cdot A}=E'$ while females have $\xi_{i\cdot A}=1-E' \ (i \in \{N,B\})$, so that male condition is better at higher $E'$ values while female condition is better at lower $E'$ values.

For males, condition affects fitness by influencing the probability of mating, assuming that at least one of the females in the patch is choosy.
We model the relative preference of a female for a male of condition $\xi$ using a beta distribution function $\mathcal B(\xi; \alpha, \beta)=\Gamma(\alpha)\Gamma(\beta)/\Gamma(\alpha+\beta)$, where $\xi\in(0,1)$, and $\alpha$ and $\beta$ are female preference factors that determine the shape of the preference curve.
Random mating occurs when $\alpha=\beta=1$.
Increasing $\alpha$ makes females decrease their preference for low-condition males but increase their preference for high-condition males; 
increasing $\beta$ makes females decrease their preference for high-condition males but increase their preference for low-condition males.
Some pairs of $\alpha$ and $\beta$ values yield a hump-shaped preference for males in intermediate condition, and others give a U-shaped preference for males with either low or high condition. 
In section B of the ESM we explain our motivation for choosing this flexible function.

For females, $\xi$ affects the number of offspring produced. 
In simulations assuming global female competition, the number of offspring produced by a females with condition $\xi$ is drawn from a Poisson distribution with mean $5\xi$ (i.e. selection on female condition is 'hard').
To keep population size constant, we randomly cull each offspring cohort to 5000.
If females compete locally (i.e. selection of female condition is 'soft'), female fecundity is normalised so that a female with condition $\xi_j$ in a patch of $n_f$ females produces $5000\xi_j/K_e\sum_i^{n_f}\xi_i$ offspring, rounded to the nearest integer, where $K_e$ is the number of patches containing at least one male and one female.

We assume that females mate with a single male, though males can mate multiply.
Each female chooses a mate from her breeding patch;
each male's probability of being picked depends on the female's preference function and the male's condition relative to his competitors (i.e. selection on male condition is always soft).
Each offspring inherits one allele from its father or mother with equal, independent probability at each of the six loci.
Mutations occur by varying the allelic values at each locus by a normally-distributed random number with mean 0 and standard deviation $\mu$, where $\mu=0.001$ for the female preference loci and $\mu=0.01$ for the other four loci.
We use a smaller mutation size at the mate choice loci because the shape of female preference function is very sensitive to the allelic values, so larger $\mu$ values make the simulations less repeatable.
The sex of each offspring is determined randomly (1:1 sex ratio).

\subsubsection*{Dispersal}
Each individual carries two loci controlling sex-specific dispersal, termed $d_f$ (only expressed in females) and $d_m$ (only expressed in males).
Each individual disperses with a probability equal to its allelic value at the relevant locus.
Those that disperse move left or right (with equality probability) a certain number of patches, where the probability of moving $x$ patches is $p(1-p)^{x-1}$ (consider a null model for random walk dispersal with constant settlement probability; see \citep{paradis:EM:2002}), with $p=0.35$ in all simulations.
We assume that passing through each patch is equally risky, such that the mortality rate of an individual that crosses $x$ patches during dispersal follows a cumulative geometric distribution function $1-(1-p_m)^x$, where $p_m=0.1$ in all simulations.

\subsubsection*{Running the simulations}
We initialised the metapopulation by randomly distributing 5000 'founder' individuals with a 1:1 sex ratio across 50 habitat patches.
In the founders, the local adaptation and sexual conflict loci take allelic values uniformly distributed between 0 and 1. 
In some simulations (indicated in the \textbf{\textit{Results}}), we fixed the dispersal probability loci $d_m$ and $d_f$ at constant values, while in others we set $d_m=d_f=0.15$ in the founders and then allow both loci to evolve.
The loci controlling the female preference function, $\alpha$ and $\beta$, were set to 1 in all founders (i.e. random mate choice).

After dispersal and before reproduction, we recorded the mean dispersal probabilities, the mean condition within each sex, the mean $\alpha$ and $\beta$ values of females, and the allelic value distribution at the mate choice and sexual conflict loci.
After the offspring generation replaced their parents, we updated the environmental state $E$ (and $E'$ if sexual conflict arises from sex-specific environmental optima) of each patch (except in simulations with a static environment), and began the life cycle anew.
Each simulation was run for $10^4$ generations.
By the end of the simulations, the conditions of males and females, the sexually antagonistic trait distribution, and the sex-specific dispersal probability (if allowed to evolve) had long reached equilibrium; the $\alpha$ and $\beta$ values at the female preference loci were sometimes still variable, but within the range that the fluctuations do not influence the shape of the female  preference function qualitatively.

\section*{Results}
\subsubsection*{IASC alone can drive the evolution of female choice in constant and homogeneous environment when females are under harder selection than males}
\begin{figure}[!h]
\begin{center}
\includegraphics[width=1\textwidth]{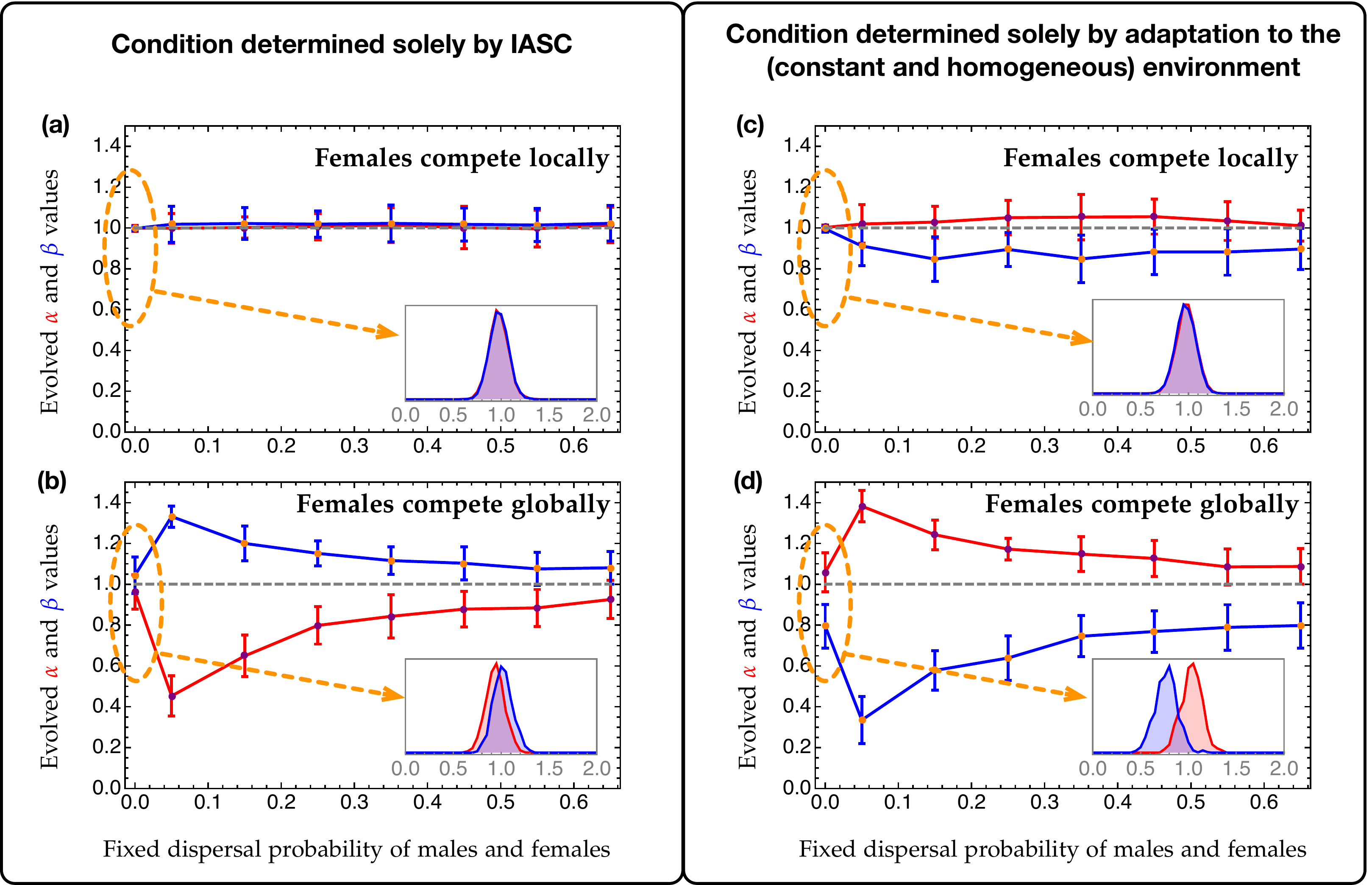}
\end{center}
\caption{Evolved $\alpha$ (red) and $\beta$ (blue) values under constant and homogeneous environment as dispersal probability increases, when condition is determined either solely by IASC or solely by adaptation to the environment, and when females compete either locally (soft selection) or globally (hard selection). The error bar plots show the mean and standard deviation of 30 independent simulation realisations. The inset in each panel shows the distribution of $\alpha$ (red) and $\beta$ (blue) values among individuals within a metapopulation at the end of simulation.}
\label{fig:Figure1}
\end{figure}

Under a constant and homogeneous environment across habitat patches, when individual condition is determined solely by the IASC locus and both sexes compete locally (i.e. soft selection), 
we find that female choice does not evolve at any dispersal rate ($\alpha\approx\beta\approx 1$,  Figure \ref{fig:Figure1}a). 
When dispersal is absent (inset of \ref{fig:Figure1}a), we recover a result from the dispersal-free model of Albert and Otto (\citep{albert:Science:2005}; model assuming autosomal linkage).

In contrast, under global competition between females (i.e. hard selection on females, soft selection on males), females evolve to prefer low-condition males ($\alpha<1$ and $\beta>1$), particularly when the dispersal rate is non-zero (Figure \ref{fig:Figure1}b).
This result occurs because females choosing a low-condition male benefit by producing fitter daughters due to IASC, which is especially important when selection on condition is harder in females than males.
Conversely, choosing a male that will produce high-condition daughters is less important under local female competition, because the resulting soft selection causes the relationship between condition and fitness to saturate faster (since attaining high enough condition to monopolise productivity in the local patch is easier than for the whole metapopulation).

One might think that the fitness costs of producing low-condition sons would be considerable, discouraging a preference for low-condition males, but it is important to note that, we assume that a male's condition only affects his fitness by influencing his attractiveness to females. 
When most females prefer low-condition males, alleles that reduce male condition actually confer a direct fitness benefit to the males carrying them.
The surprising result that females evolve to preferably mate with low-condition males was also found in previous models (e.g. \citep{albert:Science:2005,seger:Nature:1986}).
The result implies that the invasion of a new female preference strategy favouring male phenotypes that are unattractive to the resident strategy can proceed as an evolutionary runaway.
Specifically, the costs of this strategy (i.e. sons that are unattractive to females playing other strategies) diminish as the preference for low-condition males becomes more common.

If individual condition is determined solely by local adaptation, the only case where female choice cannot evolve is the combination of soft selection and the absence of dispersal (the inset of Figure \ref{fig:Figure1}c); otherwise, females evolve to prefer high-condition males.
This result is intuitive because high-condition males are more locally-adapted and there is no reduction in daughter fitness since IASC is absent.
Females who prefer to mate with high-condition males will produce better-adapted sons that will be preferred by other choosy females in the population.

When selection on females is hard, female preferences are strongest when dispersal is present but rare; high dispersal rates result in weaker preference, which was also found previously \citep{veen:JEB:2015}.
When selection is soft on females, dispersal rate (when positive) has little effect on the strength of female preference.

\subsubsection*{Coevolution of female choice and dispersal}

\begin{figure}[!h]
\begin{center}
\includegraphics[width=1\textwidth]{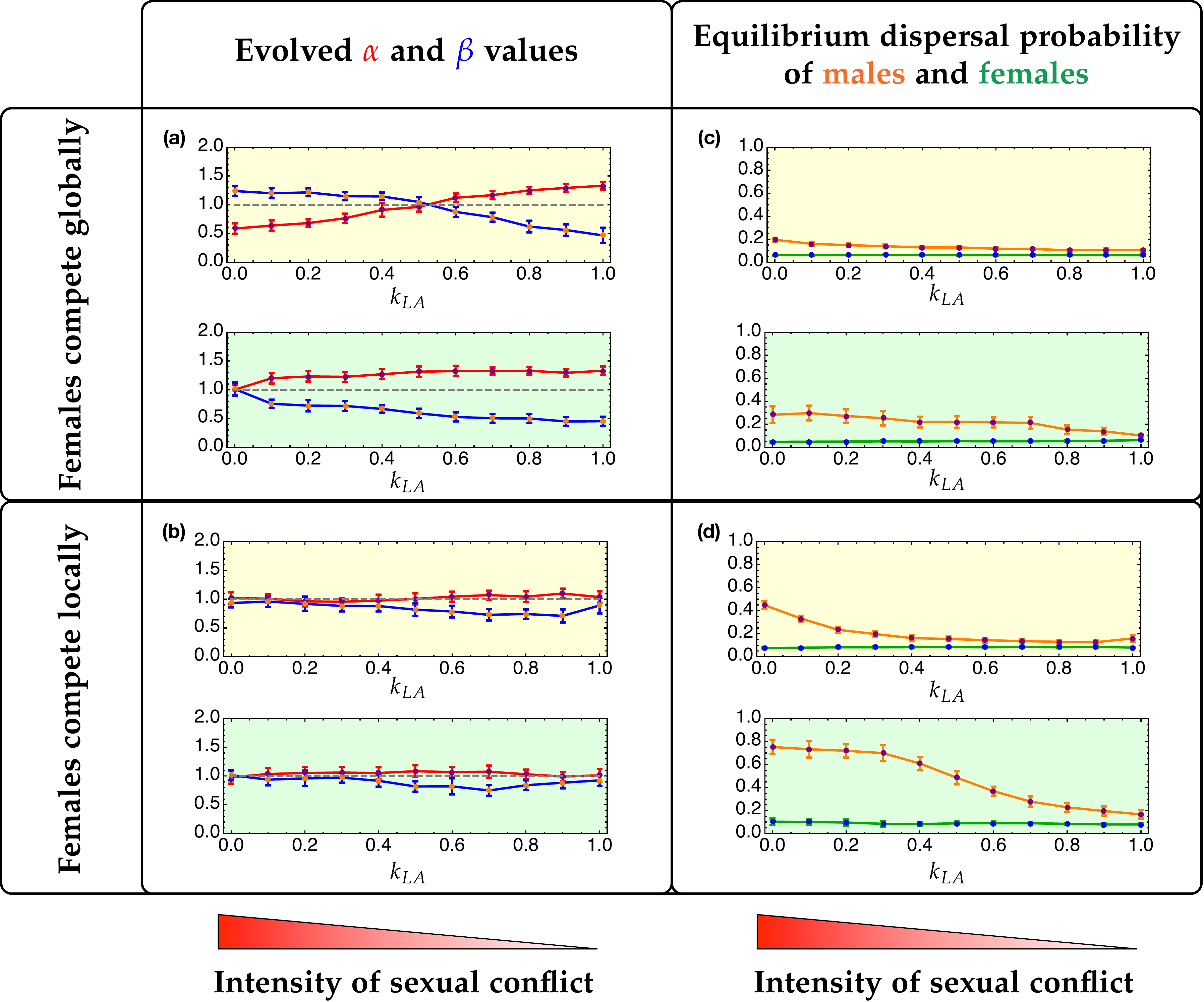}
\end{center}
\caption{Evolved $\alpha$ (red) and $\beta$ (blue) values and equilibrium dispersal probabilities of males (orange) and females (green) under different intensity of sexual conflict, when females compete either globally or locally. Panels with yellow background represent the cases where sexual conflict arises from IASC, and panels with green background represent the cases where males and females have different environmental optima. The error bar plots show the mean and standard deviation of 30 independent simulation realisations.}
\label{fig:Figure2}
\end{figure}

In the previous section, we assumed that the dispersal probabilities of males and females were fixed, and we also set $k_{LA}$ (the relative weight of local adaptation in determining individual condition) either to 0 (only IASC is present) or 1 (only local adaptation is present).
Now we vary the parameter $k_{LA}$ from 0 to 1, and allow the sex-specific dispersal probabilities to coevolve with female choice (Figure \ref{fig:Figure2}).
Next to a scenario where sexual conflict is caused by the IASC locus (yellow panels in Figure \ref{fig:Figure2}), we also study a scenario where sexual conflict results from sex differences in environmental optima (green panels in Figure \ref{fig:Figure2}).
Note that the two scenarios become identical at the boundary case where $k_{LA}=1$.
Consistent with the previous section, the environment is still kept constant and homogeneous across habitat patches.

We find that when selection on females is hard and sexual conflict arises from IASC (Figure \ref{fig:Figure2}a, yellow panel), females evolve to prefer low-condition males if sexual conflict is strong ($k_{LA}$ is relatively small), but prefer high-condition males when local adaptation becomes more important ($k_{LA}$ is relatively large).
In contrast, if sexual conflict arises from a difference in environmental optima between the sexes, females always evolve to prefer high-condition males, and the preference increases with the relative weight of local adaptation (Figure \ref{fig:Figure2}a, green panel).
In this case, since all males respond to the constant environmental condition in the same way, the differences in their condition candidly reflect the differences in their degree of local adaptation.

When selection is soft on both sexes (Figure \ref{fig:Figure2}b), females never evolve to prefer low-condition males, which is consistent with our results in the previous section.
In addition, female preference for high-condition males only evolves when the relative weight of local adaptation in determining individual condition ($k_{LA}$) is relatively high, no matter whether sexual conflict arises from IASC or different environmental optima between the sexes.
Interestingly, the strength of female choice first increases and then decreases as $k_{LA}$ approaches 1 (Figure \ref{fig:Figure2}b). 
This result is consistent with our prediction that the presence of a moderate amount of sexual antagonistic variance in fitness increases the fitness benefits of female choice, e.g. by causing females to occasionally mistake male-beneficial genotypes for locally-adapted ones.

Starting from the same initial dispersal probability ($d_m=d_f=0.15$), females always evolve to an equilibrium dispersal probability close to 0 (but not equal to 0, due to mutation-selection balance), and males evolve higher dispersal probabilities than females.
%One reason for this is the polygynous mating system, where kin competition for mating happens between males in the same breeding patch, while females experience no mating competition with each other.
The combined effect of polygynous mating system and demographic stochasticity causes spatiotemporal variation in the reproductive success to be higher for males than for females \citep{gros:TPB:2009, henry:AmNat:2016}.
Additionally, male dispersal probability decreases as $k_{LA}$ increases;
this is because local adaptation is more important to fitness, and condition provides more information about male genetic quality as $k_{LA}$ increases.

In addition, dispersal is more male-biased when females compete locally rather than globally, especially when $k_{LA}$ is small.
This is because the odds of dispersing to a patch containing many females are higher when females are under soft selection (i.e. all patches produce the same amount of offspring) rather than hard selection (causing some patches to be empty).
Under global female competition, the number of offspring produced per patch can differ greatly, resulting in "super-patches" containing many individuals, and because most successful dispersal events are relatively short-ranged, the typical meta-population state is that most individuals are located in a few clusters of highly populous patches separated by empty space (this was determined by inspecting several simulations). 
Therefore, compared to local competition, when females are under global competition, the patch in which a male is born is more likely to contain more females than the one he migrates to, making it worth staying despite the presence of more competitors.

Furthermore, the equilibrium dispersal probability of males is higher when sexual antagonism arises from a sex difference in environmental optima than when it arises from IASC, especially when $k_{LA}$ is small (Figure \ref{fig:Figure2}c-d, compare the green panels and yellow panels).
When sexual conflict arises from sexually dimorphic environmental optima, all males in the same patch react to the environment in the same way, and thus have exactly the same sexually antagonistic component of condition, and consequently, the overall conditions of males are more similar than the case where sexual conflict arises from IASC.
The small differences between male conditions reduce the benefit of female choice, and the consequent random choice of females intensifies kin competition between brothers, selecting for higher dispersal.

\subsubsection*{Effect of environmental heterogeneity and stochasticity}
\begin{figure}[!h]
\begin{center}
\includegraphics[width=1\textwidth]{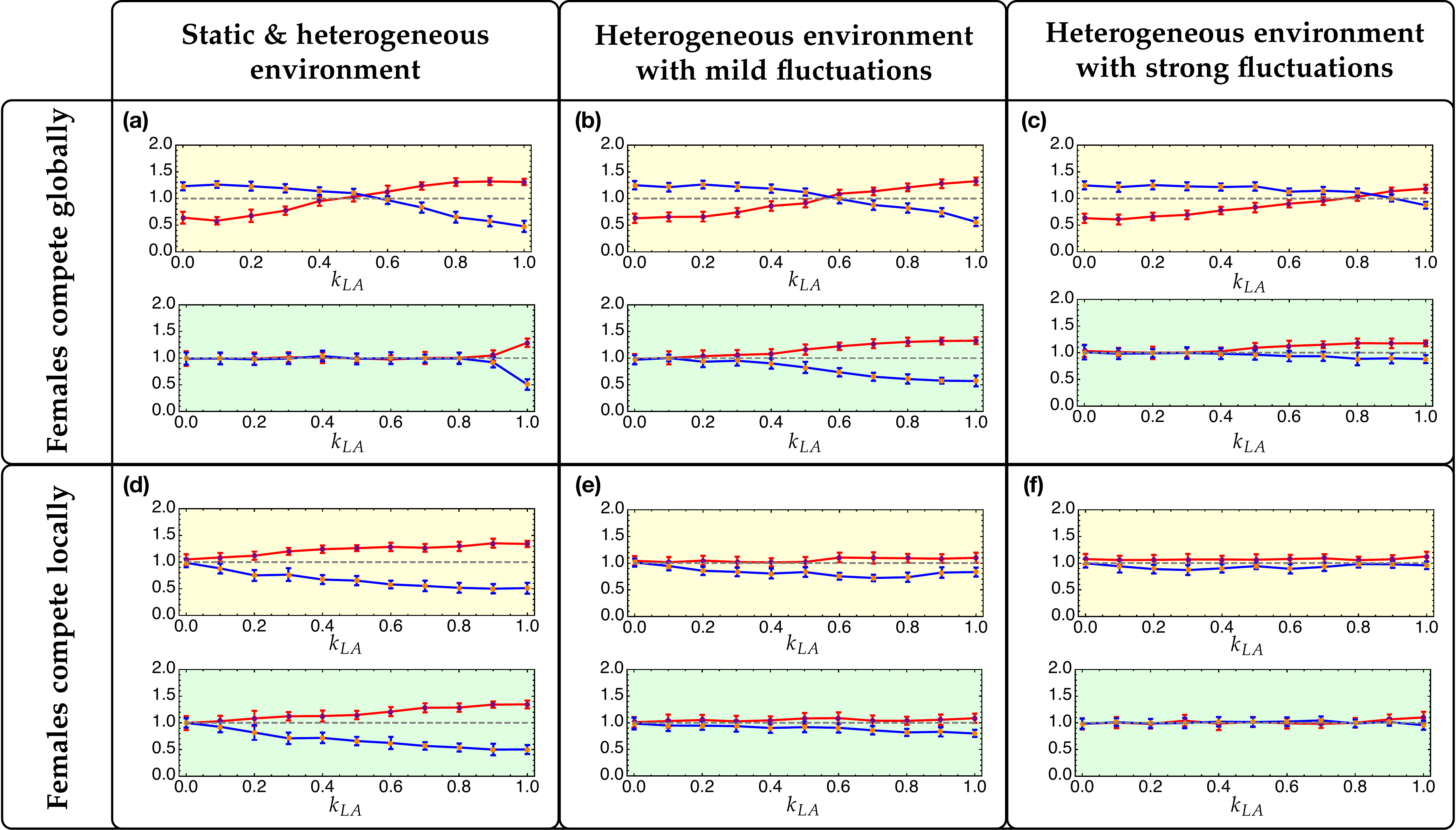}
\end{center}
\caption{Evolved $\alpha$ (red) and $\beta$ (blue) values when female choice coevolves with sex-specific dispersal probability. The arrangement of panels in each column is the same as in the first column of Figure \ref{fig:Figure2}. Each data point in the error bar plots represents the mean and standard deviation from 30 independent simulation realisations. }
\label{fig:Figure3}
\end{figure}

Up to now we have always kept the environmental conditions constant and homogeneous; we now relax these assumptions and allow the environment to vary across patches and fluctuate over time. 
As shown in Figure \ref{fig:Figure3}, some general patterns still hold, for example: 
1) females can evolve to prefer low-condition males only if sexual conflict arises from IASC, selection is hard on females, and $k_{LA}$ is relatively small;
2) females always evolve to prefer high-condition males when sexual conflict is weak and condition depends mainly on the degree of local adaptation ($k_{LA}$ is close to 1).

The different types of environments also have their specific features.
For example, when the environment is static and heterogeneous and sexual conflict arises from different environmental optima between the sexes, female preference for high-condition males easily evolves when females compete locally, but not so easily when females compete globally (compare the green panels in Figure \ref{fig:Figure3}a and \ref{fig:Figure3}d).
This pattern is the opposite to the previous case of environmental homogeneity (green panels in Figure \ref{fig:Figure2}a and \ref{fig:Figure2}b). 
This is because environmental heterogeneity allows for local adaptation, such that females benefit by avoiding maladapted immigrant males \citep{kokko:genetica:2008};
the effect is weaker under global competition since then, most individuals live in a few ``super patches'' to which they are already well adapted.
When the environment also fluctuate over time, the parameter space where females prefer high-condition males shrinks and the strength of female preference also weakens as the environmental fluctuation becomes stronger, as the benefit from local adaptation lessens.

\subsubsection*{Effect of female choice on the condition of males and females and IASC allelic value distributions}

\begin{figure}[!h]
\begin{center}
\includegraphics[width=1\textwidth]{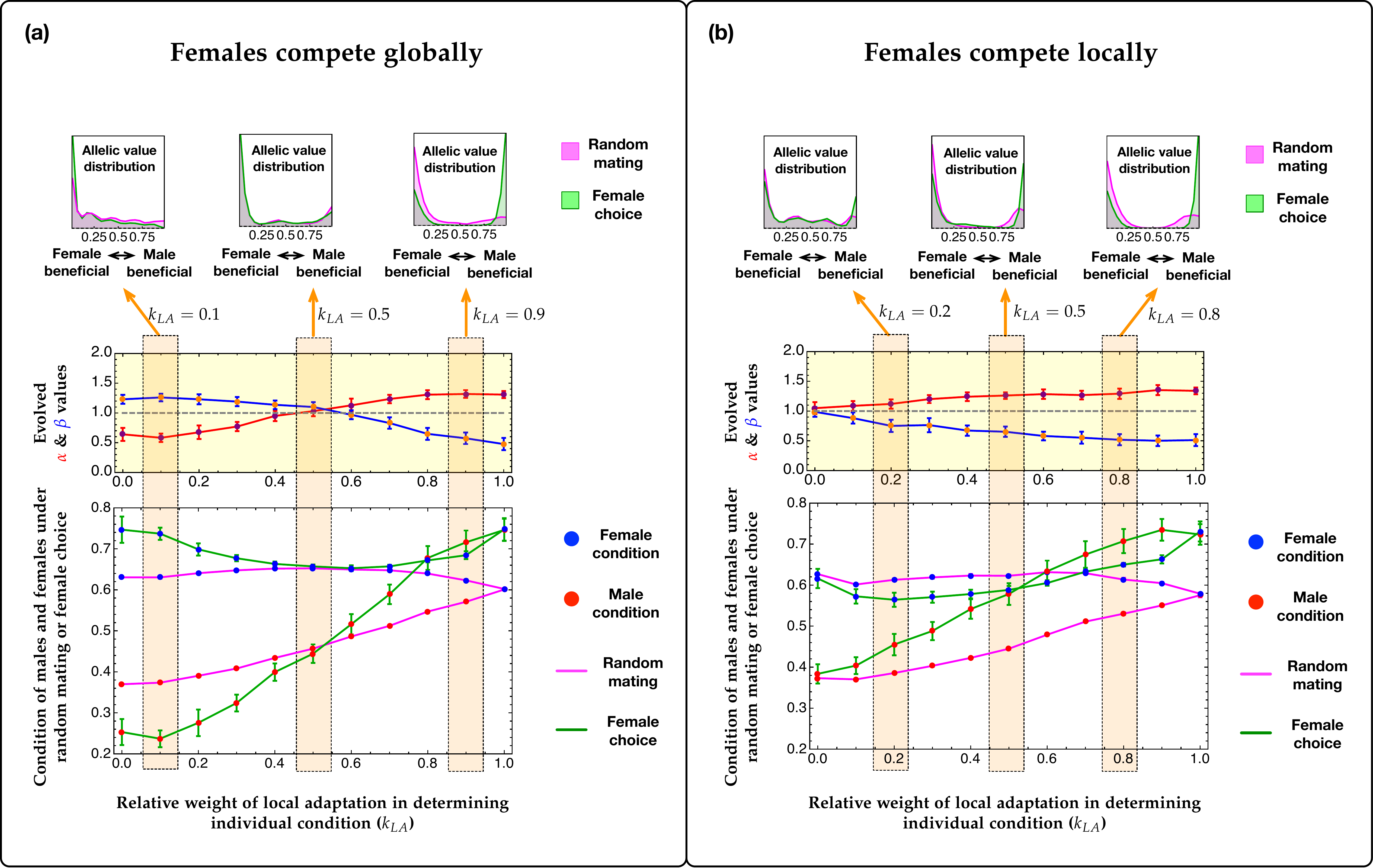}
\end{center}
\caption{Equilibrium condition of males and females and allelic value distributions at the IASC locus when female choice is either allowed or prevented intentionally, when females compete either globally or locally. The environment is static and heterogeneous across habitat patches. Each data point in the error plots represents the mean and standard deviation of 30 independent simulations. The allelic distributions of the IASC locus is the state at the end of the simulation ($10^4$ generations), also the mean from 30 independent simulations.}
\label{fig:Figure4}
\end{figure}

To understand the effect of female choice on the condition of males and females, we created scenarios where the evolution of female choice was prevented.
In Figure \ref{fig:Figure4}, we show the equilibrium conditions of males and females and the allelic value distribution at the IASC locus when female choice is either allowed to evolve or prevented, when females compete globally or locally.
The environment is set to be static and heterogeneous across habitats in Figure \ref{fig:Figure4}; 
results under the other environmental conditions are qualitatively similar (see ESM section C).

To derive a baseline expectation for the equilibrium condition of males and females, we purposely prevented female choice from evolving by setting $\alpha=\beta=1$ and preventing mutation at the female choice loci.
When female choice naturally cannot evolve (without being prevented), the equilibrium condition of males and females should be identical to the corresponding baseline cases, such when $k_{LA}=0.5$ under global female competition and when $k_{LA}=0$ under local female competition, as shown Figure \ref{fig:Figure4}.

When females compete globally, they always have either higher or equal condition when female choice is allowed than when females are forced to mate randomly.
When females evolve to prefer low-condition males (e.g. when $k_{LA}=0.1$ in Figure \ref{fig:Figure4}a), females have higher condition because the low-condition males carry female-beneficial alleles at the IASC locus, as shown in the allelic value distribution, the frequency of female-beneficial alleles (with allelic values close to 0) is higher when female choice is allowed (green) than when females are forced to mate randomly (magenta).
When females evolve to prefer high-condition males (e.g. when $k_{LA}=0.9$ in Figure \ref{fig:Figure4}a), choosy females are better off in condition than random-mating females because they benefit from mating with locally adapted males.
Note that in this parameter region, $k_{LA}$ is high and sexual conflict is weak, so that the benefit from mating with high-condition (likely locally-adapted) males offsets the detrimental effects of IASC on daughter fitness.
As shown in the allelic value distribution, male-beneficial alleles (with allelic values close to 1) have higher frequency when female choice is allowed.

In contrast, when females compete locally so that each of the habitat patch produces the same number of offspring, female conditions sometimes are lower when mate choice is allowed than when it is prevented on purpose (e.g. when $k_{LA}=0.2$ in Figure \ref{fig:Figure4}b), suggesting a 'tragedy of the commons' whereby the evolved female mating strategy (i.e. 'accept some reduction in daughter fecundity in exchange for elevated son mating success') reduces the average number of offspring produced by each individual in the population.
The tragedy of the commons happens when sexual conflict is strong, and disappears when the relative weight of local adaptation in determining individual condition ($k_{LA}$) becomes large enough (see results for alternative environmental conditions in Figure S5 of the ESM).
When females are under soft selection, the evolution of female choice always shifts the distribution of allelic values towards male-beneficial and female-detrimental alleles, particularly when $k_{LA}$ is large (meaning IASC has only a small/moderate effect on condition).
Additionally, the evolution of female choice creates a bimodal distribution of allelic values at the IASC locus, with the frequency of male-beneficial alleles rises as $k_{LA}$ increases, showing that the evolution of female choice can elevate the proportion of genetic variance in fitness that is sexually antagonistic, which concurs with the results of \citep{arnqvist:evolution:2011}.

\section*{Discussion}
Mate choice evolution is a complex problem when one jointly considers sexually concordant and antagonistic variance in fitness, the scale of competition, and coevolution between mate choice, adaptation, and dispersal.
Although we recapitulated the results of Albert and Otto \citep{albert:Science:2005} that females should not evolve mate choice for phenotypes encoded by sexually antagonistic autosomal alleles (Figure \ref{fig:Figure1}a inset), 
we find that female choice does evolve when individual condition jointly depends on IASC and local adaptation.
The strength and even the sign of the female preference can change, depending on the relative importance of IASC and local adaptation, which in turn is likely to depend on multiple evolving and non-evolving parameters (e.g. the demography and evolutionary history of the population, and spatiotemporal variance in the abiotic environment).

We found some evidence that IASC can select for stronger female choice by helping to preserve genetic variation, as previously found for GEIs \citep{kokko:genetica:2008,holman:chapter:2014}, though the effect was weak.
Rather, female preference for high-condition males weakens as IASC intensifies (i.e. with decreasing $k_{LA}$), because the benefits of getting locally-adapted alleles and fit sons are partly negated by reduced daughter fitness.
Female preference for high-condition males diminished, and sometimes even flipped towards low-condition males, when selection to maximise daughter fitness is sufficiently strong.
This surprising result only occurs when females compete globally (i.e. when female condition is under hard selection), making it especially advantageous to produce fit daughters, and when environmental turnover is rapid, making it less important to obtain locally-adapted alleles.

Previous theoretical work has provided additional reasons why females might evolve to prefer low-condition males, such as sex linkage of female choice and/or sexual conflict locus \citep{albert:Science:2005}, and parental care \citep{cotar:JTB:2008}.
It is also shown experimentally that zebra finches (\textit{Taeniopygia guttata}) females prefer high/low-condition males if they themselves are of high/low condition, perhaps because low-condition females benefit by avoiding competition with high-condition females for the best males \citep{holveck:PBS:2010}.
In the context of \citep{cotar:JTB:2008, holveck:PBS:2010}, mate competition among females was a prerequisite for some females to prefer low-condition males. 
Here, we showed that IASC can cause females to prefer low-condition males even when female mate competition is absent.

Our model showed that when selection on females is hard (global) while selection on male is soft (local), females are more likely to evolve a preference for female-beneficial, male-detrimental genotypes.
In natural populations of animals, selection is often softer on males than on females, because soft selection results from local density- and frequency-dependent effects on fitness \citep{reznick:JH:2016}, and male fitness is often more strongly affected by the number and quality of local same-sex competitors than is female fitness \cite{bateman:heredity:1948,janicke:SA:2016}.
This trend suggests that IASC probably does limit the evolution of female preferences for 'male' phenotypes (i.e. those detrimental when expressed in females) in most species.
However, the magnitude (and perhaps sign) of the sex difference in the softness of selection may vary substantially between taxa; for example, it might be comparatively small in species/populations with strict monogamy, obligate paternal care, or intense local female-female competition (all of which affect the relative scale of competition in each sex).

To illustrate the application of this idea, we consider two \textit{Drosophila} experimental evolution studies, both of which manipulated the mating system and then tested whether the transcriptome evolved masculinisation or feminisation, but which obtained opposite results.
Our model suggests a novel explanation for this disparity.
Hollis \textit{et al.} (2014) \citep{hollis:NatComm:2014} and Veltsos \textit{et al.} (2017) \citep{veltsos:NatCommun:2017} both experimentally imposed random monogamy by rearing flies in vials ('subpopulations') containing 1 male and 1 female, and predicted that monogamy would remove the benefits of male competitive adaptation, and thereby select for transcriptional feminisation.
However, the studies' non-monogamous treatments were different:
polygynandry (5 males and 5 females per subpopulation) was used in \citep{hollis:NatComm:2014}, while polyandry (6 males and 1 female) was used in \citep{veltsos:NatCommun:2017}. 
The polygynandry treatment facilitates local female competition, while the polyandry treatment imposes global selection on females.
This means that in  \citep{hollis:NatComm:2014}, the monogamy treatment removes sexual selection on males as well as hardening selection on females (relative to the polygynandrous control), both of which are predicted to feminise the phenotype, and was observed in  \citep{hollis:NatComm:2014}.
By contrast, hard selection on female condition is imposed under both treatments in \citep{veltsos:NatCommun:2017}.
Our model predicts that mate choice feminises the phenotype under hard selection (compare Figure \ref{fig:Figure4}a and \ref{fig:Figure4}b; the condition of females/males is generally higher/lower under global female competition), so removing choice via enforced monogamy should masculinise the phenotype, as found in \citep{veltsos:NatCommun:2017}.
Our explanation is not mutually exclusive with other hypotheses for the discrepancy in results (see \citep{veltsos:NatCommun:2017}), but it highlights the value of considering the sex-specific softness of selection, and perhaps addressing it with targeted experiments, in future work.

Another interesting result is that the evolution of female mate choice sometimes caused a 'tragedy of the commons' when females are under soft selection and IASC is strong, whereby females evolved to prefer males carrying alleles that reduce fecundity of the female's daughters, causing a population-wide fitness reduction relative to a randomly-mating population.
Female choice also preserved genetic variation at the antagonistic locus, exacerbating IASC.
This result echoes recent empirical work in the seed beetle \textit{Callosobruchus maculatus} showing that selection favours genotypes that reduce female fitness and population productivity, because these genotypes pleiotropically elevate male mating success \citep{berger:AmNat:2016}. 
Furthermore, this result has conservation implications.
Selective harvesting of males with exaggerated secondary sexual traits has been proposed to greatly harm population mean fitness by removing highly attractive males carrying 'good genes' \citep{knell:PRSB:2017}, but we hypothesise that IASC would weaken or reverse this effect.
Similarly, captive breeding programs that allow animals to choose their mates in order to aim to maximise genetic quality \citep{wedekind:CB:2002} may end up favouring male-beneficial, female-detrimental alleles; it might be better to breed from males who have fit female relatives.
Lastly, anthropogenic change that disrupts female choice has been suggested to genetically weaken population \citep{kempenaers:CB:2010, sundin:Ethology:2010}, but again, this assumes that females prefer genotypes that elevate population fitness.

In this work, we limited the types of sexual conflict to IASC and sexually dimorphic environmental optima, but it is worth noting that IASC and \textit{inter}-locus sexual conflict (IRSC) are closely linked \cite{bonduriansky:TREE:2009,perry:CSH:2015}.
Resolving IASC by allowing each sex to reach its own phenotypic optimum does not necessarily improve the fitness of the population, particular because 'well-adapted' males may possess more harmful competitive adaptation (that is, reducing IASC might elevate IRSC).
For example, several insect experimental evolution studies have concluded that selection favours harmful male phenotypes, reducing the fitness of females interacting with them \citep{holland:PNAS:1999,martin:BioOne:2003,gay:Evolution:2011}.
In plants, a recent experimental evolution study concluded that pollen competition selects for more competitive pollen tube growth, but these competitive male traits harm the fitness of the recipient plant (IRSC) as well as reduce seed production of the same plant (IASC) \citep{lankinen:Evolution:2017}.
Lastly, a recent model found that the interactions of IASC and IRSC can prevent populations from reaching evolutionary equilibria when female choice is under strong pleiotropic constraints, but trigger a new coevolution arms race between the sexes \citep{pennell:PNAS:2016}.
In this light, it could be interesting to extend our model to similarly incorporate IRSC, for example by reframing it as a model of male coercion and female resistance, rather than of male quality and female preference.

%\bibliography{\string~/MyLibrary/MyBib.bib}

\newpage{}
\section*{Online Supplementary Information}

\renewcommand{\theequation}{S\arabic{equation}}
% redefine the command that creates the equation number.
\renewcommand{\thetable}{S\arabic{table}}
\renewcommand{\thefigure}{S\arabic{figure}}
\setcounter{equation}{0}  % reset counter 
\setcounter{figure}{0}
\setcounter{table}{0}

\subsection*{A. Simulating different environments}

We created 4 different types of environments:

\begin{enumerate}

\item \textbf{Constant and homogeneous environment across habitat patches}:
Under this type of environment, the $E$ values in all habitat patches are set to 0.5 and do not change over time.

\item \textbf{Constant and heterogeneous environment across habitat patches}: Under this type of environment, the $E$ value in each of the habitat patches is independently drawn from a normal distribution with $\mu=0.5$ and $\sigma=0.2$, and bounded between 0 and 1. The $E$ values do not change over time. Supplementary Figure \ref{fig:FigureSIA2} shows 10 independent samples of this type of environment.

\begin{figure}[!h]
\begin{center}
\includegraphics[width=1\textwidth]{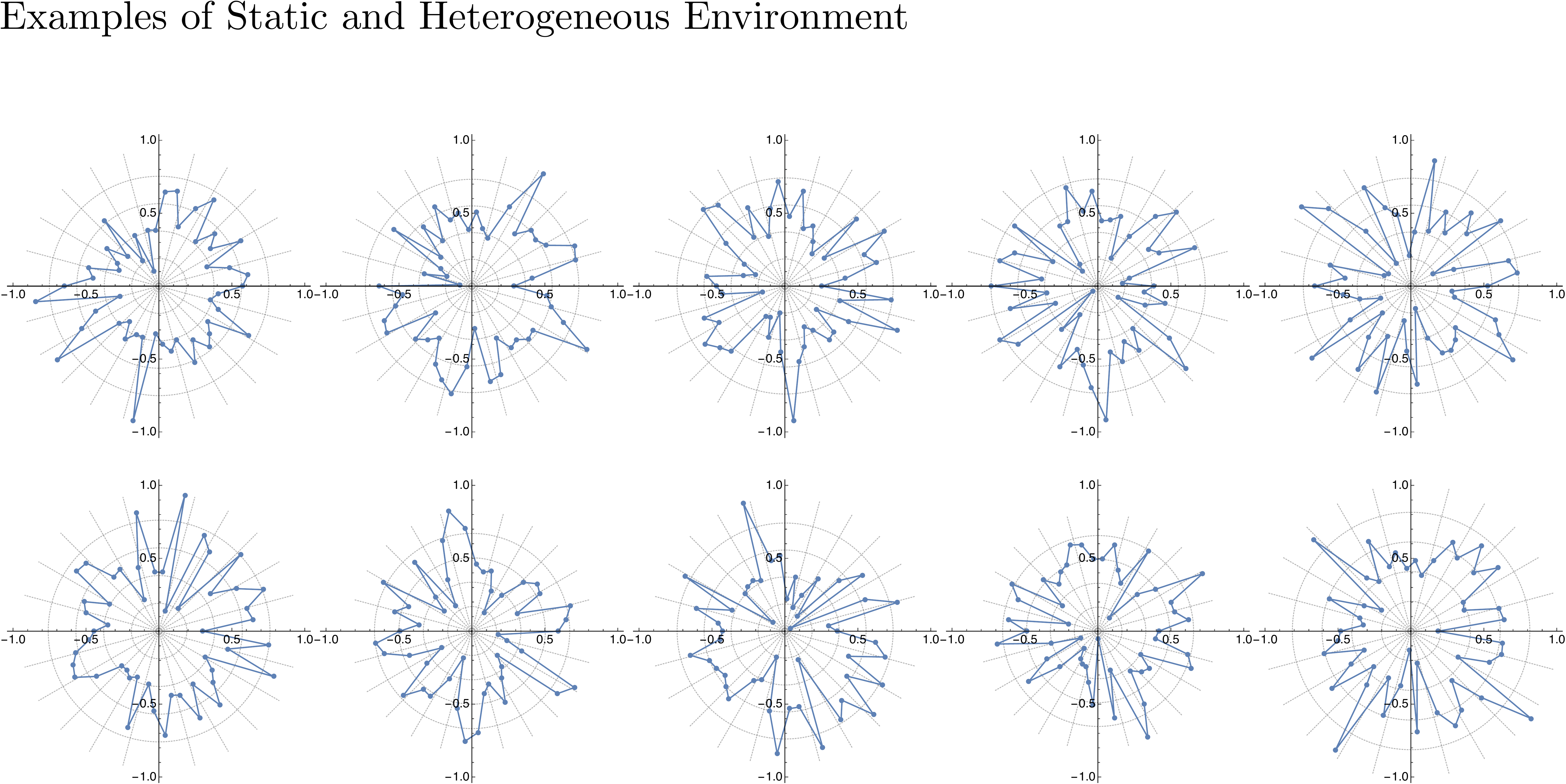}
\end{center}
\caption{Independent examples of the environmental states ($E$ values) across the 50 habitat patches (shown as radially-arranged points) under the static but spatially heterogeneous environment. }
\label{fig:FigureSIA2}
\end{figure}

\item \textbf{Heterogeneous and mildly fluctuating environment}, and
\item \textbf{Heterogeneous and strongly fluctuating environment}: When simulating these two types of environment, we assume that the $E$ values are spatially autocorrelated across space so that nearby patches tend to have similar values. 
In addition, we we allow $E$ to change over time within each patch, with temporal autocorrelation.
The environmental state $E$ on patch $j$ at generation $t$,  $E_j(t)$, is determined first by calculating its spatial component  $E_j^s(t)=p_s\langle E_j(t-1)\rangle+(1-p_s)\zeta$, in which $p_s$  ranges between 0 and 1, adjusting the degree of spatial autocorrelation,  $\langle E_j(t-1)\rangle$ is the mean environment condition on patch $j$ and its two closest neighbours at the previous generation, and $\zeta\in(0,1)$  is a uniformly-distributed random number. 
After the spatial component, temporal autocorrelation is incorporated to update  $E_j(t)$, so that  $E_j(t)=p_t E_j(t-1)+(1-p_t)E_j^s(t)$, where $p_t$  also ranges between 0 and 1, adjusting the degree of temporal autocorrelation. 

We initialise the environment by first setting $E$ to 0 or 1 with equal probability for each patch, and then letting the environment update for 500 generations, ensuring the environment has reached a dynamic equilibrium (corresponding to the specified levels of spatial and temporal autocorrelation) before introducing the population. 
We set  $p_s=p_t=0.7$ for simulating the heterogeneous and mildly fluctuating environment, and set $p_s=p_t=0.5$ for simulating the heterogeneous and strongly fluctuating environment.

Note that the method we use here for simulating spatially and temporally autocorrelated environmental fluctuation means that the spatial and temporal autocorrelations are not independent of each other; 
this method was chosen for computational efficiency (a theoretically better method would be to use a 2D Gaussian field with space and time on each axis, but this would necessitate constructing and calculating the singular value decomposition of matrices of size $(5\times10^5)\times(5\times10^5)$ for simulating the environmental states on 50 patches for $10^4$ generations, which is computationally not feasible). 
Because of this limitation, we did not vary $p_s$ and $p_t$ independently of each other, and simply examined a pair of high values and a pair of lower values. 
Supplementary Figure \ref{fig:FigureS1} shows a sample series of the environmental states across habitats through 10 consecutive generations for each of the high/low environmental fluctuation regimes. 
\bigskip\bigskip
\begin{figure}[!h]
\begin{center}
\includegraphics[width=1\textwidth]{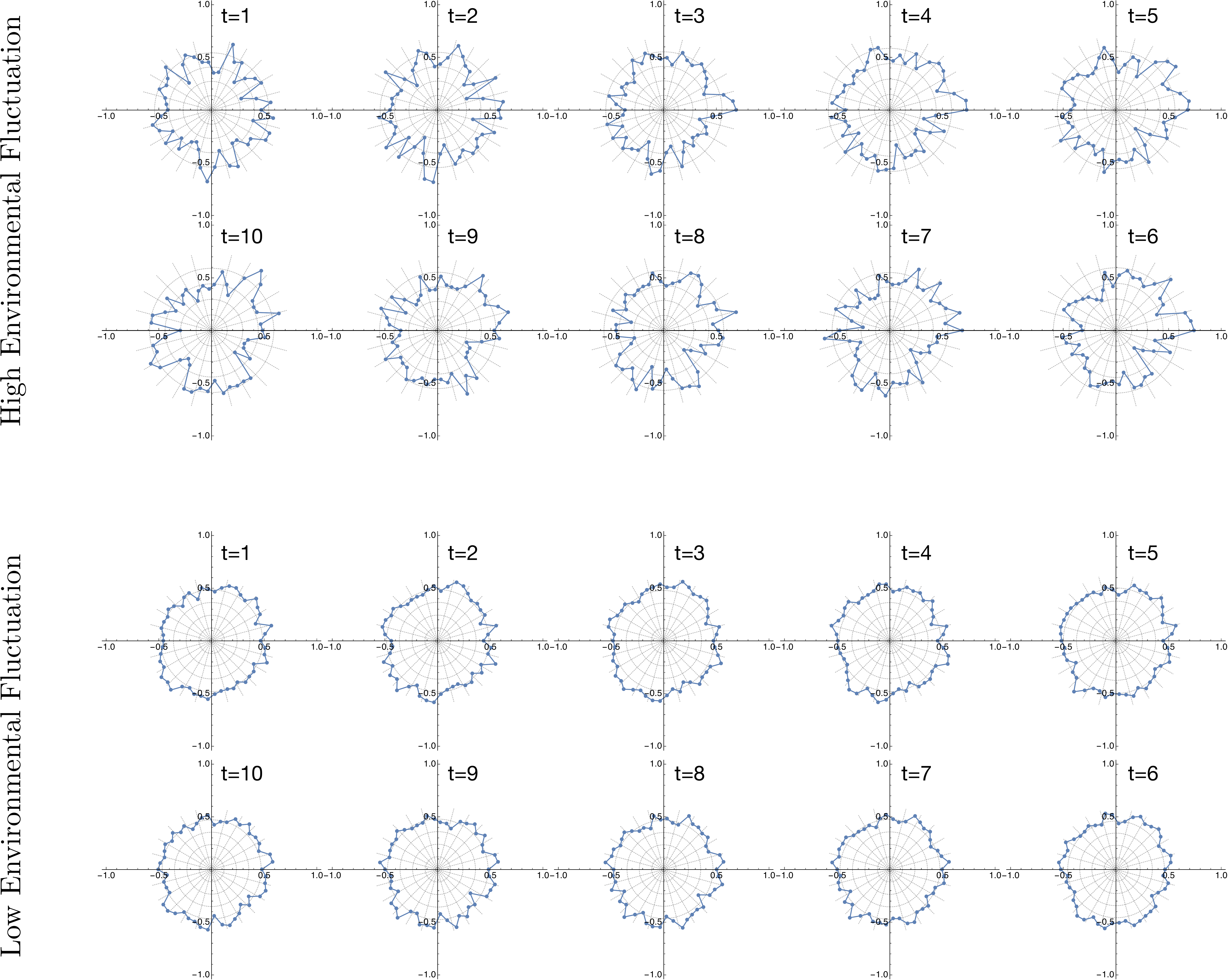}
\end{center}
\caption{The environmental states ($E$ values) of the 50 habitat patches (shown as radially-arranged points), under strong or mild environmental fluctuations ($p_s=p_t=0.7$ or $p_s=p_t=0.5$  respectively). }
\label{fig:FigureS1}
\end{figure}

\end{enumerate}

When sexual conflict arises from different environmental optima between males and females, we generate another environmental state $E'$ to determine the environmental component of individual condition. 
The $E'$ environment is of the same type (e.g. have the same degree of spatial and temporal variations) as $E$ in the corresponding simulations, but generated independently.
\bigskip\bigskip

\subsection*{B. Representing female preferences using the beta function}

Sexual selection models involving female choice typically assume that females either mate randomly or prefer mating with high-condition males if they are allowed to choose \cite{kokko:PRSB:2002, reinhold:ODE:2004, holman:chapter:2014}. 
The assumption is appropriate if the relationship between genotype and fitness is concordant between the sexes. 
But when at least some loci have sexually antagonistic effects on fitness, choosy females must trade off the fitness of sons and daughters, and it becomes harder to differentiate males that are well-adapted to the environment from males that are simply well-adapted to being male. 
With this in mind, we sought to use a more flexible female preference function than in previous models, which would also allow the evolution of preferences for low- or intermediate-condition males, if such preferences were advantageous.

\begin{figure}[H]
\begin{center}
\includegraphics[width=1\textwidth]{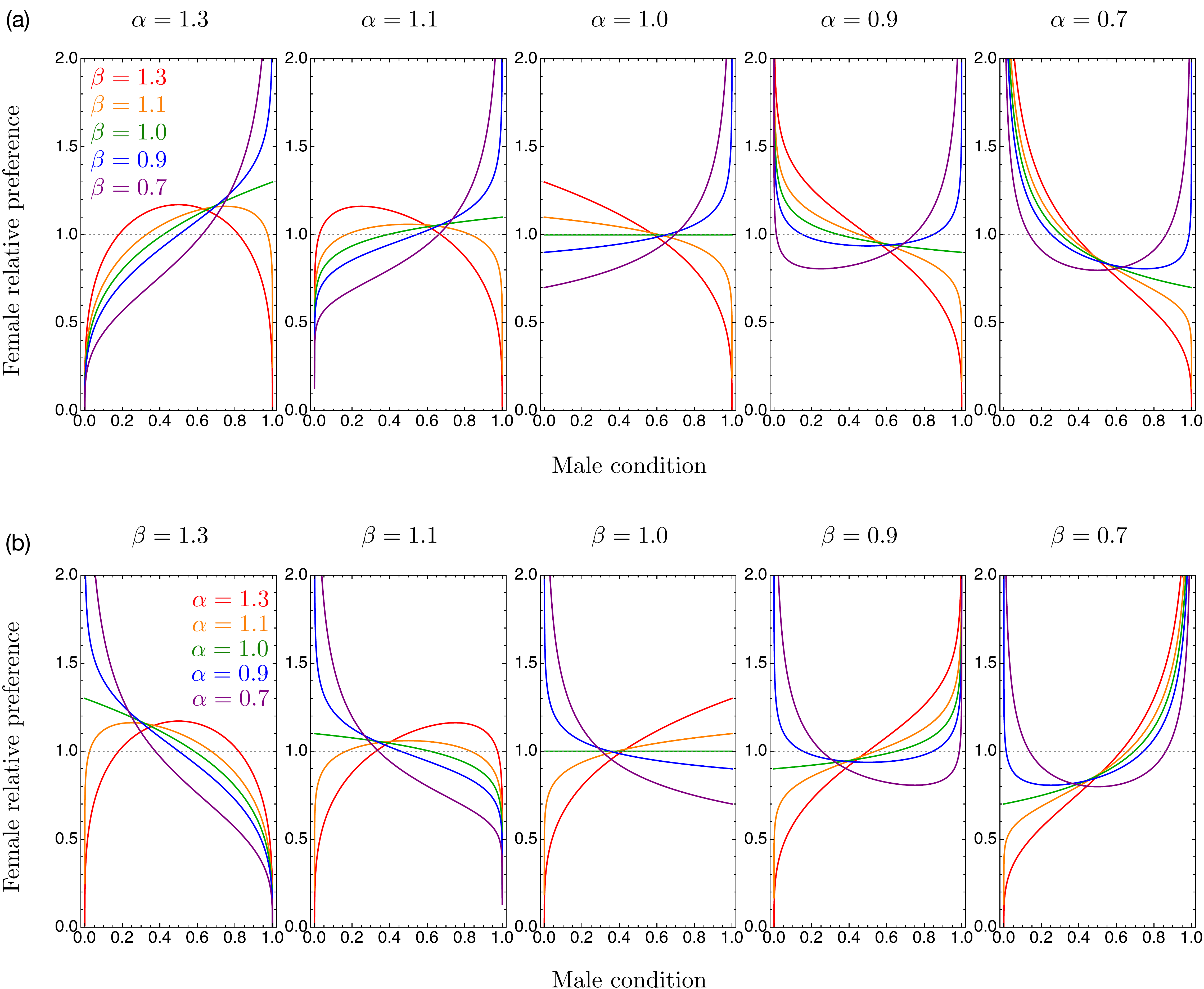}
\end{center}
\caption{Examples of female relative preference as a function of male condition $\xi$  modelled by the beta distribution function as the values of $\alpha$  and  $\beta$ vary. By allowing the loci controlling $\alpha$  and   $\beta$ to evolve, our model potentially allows the evolution of a wide range of monotonic and non-monotonic female preference functions.}
\label{fig:FigureSIB1}
\end{figure}

Therefore, we represent the preference for a male with condition  $\xi$ as  $\mathcal B(\xi; \alpha,\beta)=\Gamma(\alpha)\Gamma(\beta)/\Gamma(\alpha+\beta)$.
Supplementary Figure \ref{fig:FigureSIB1} illustrates the change of female relative preference as the two shape parameters $\alpha$ and $\beta$ vary. 
When  $\alpha$ is fixed, increasing $\beta$  causes females to increase their preference for low-condition males while decreasing preference for high-condition males; 
when  $\beta$ is fixed, increasing $\alpha$  causes females to increase preference for high-condition males and decrease preference for low-condition males. 
When  $\alpha=\beta=1$, females prefer males of all conditions equally, representing random mate choice. 

Qualitatively, the female preference function is increasing (females prefer high-condition males over low-condition males) when $\alpha \geqslant 1$ and $\beta<1$, and decreasing (females prefer low-condition males over high-condition males) when $\alpha<1$ and $\beta \geqslant 1$.

\subsection*{C. Additional results}

\subsubsection*{Evolved dispersal probability under different types of environments}

\begin{figure}[H]
\begin{center}
\includegraphics[width=1\textwidth]{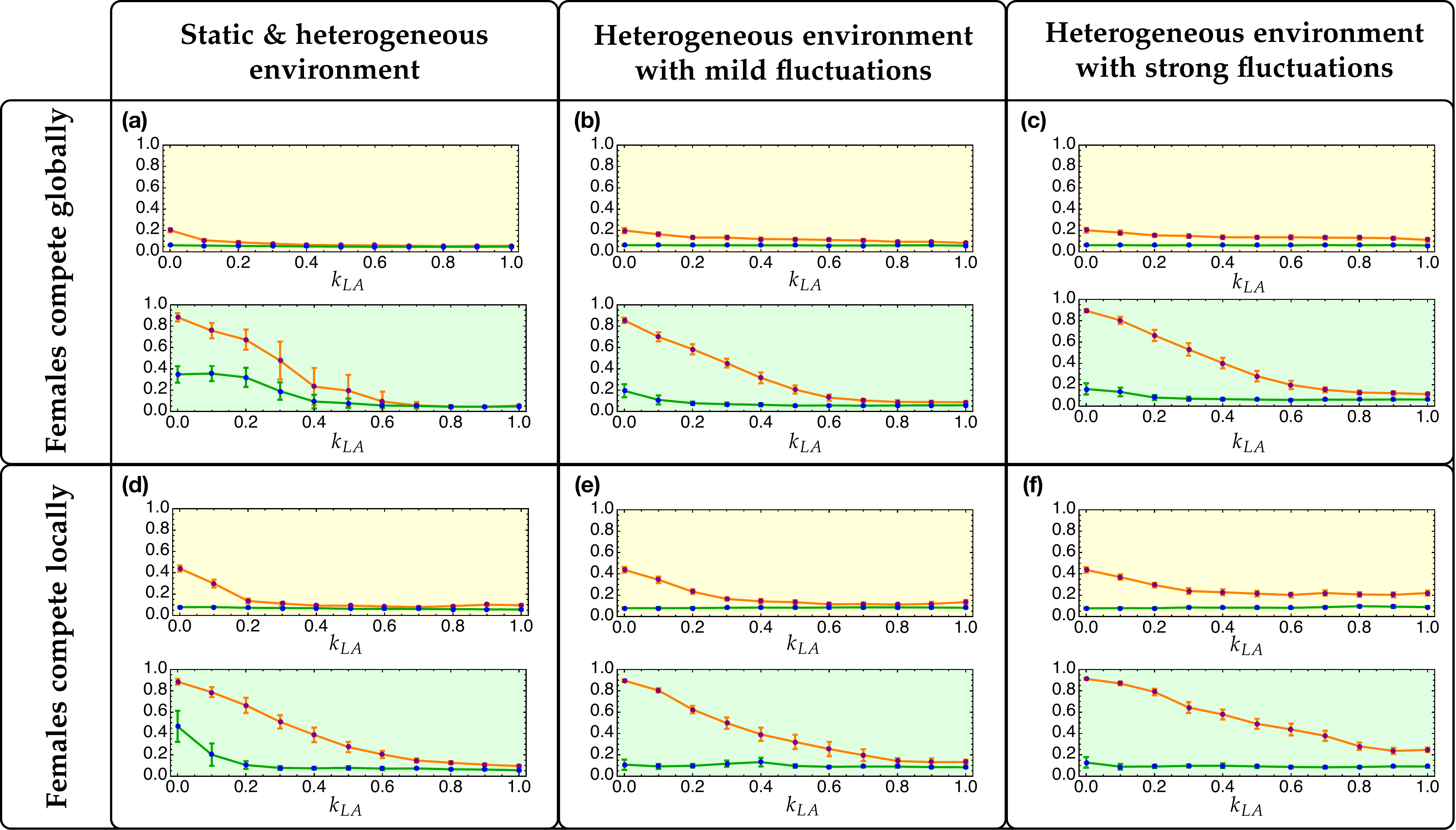}
\end{center}
\caption{Equilibrium dispersal probability of males (orange) and females (green) under different intensity of sexual conflict, when females compete either globally or locally. Panels with yellow background represent the cases where sexual conflict arises from IASC, and panels with green background represent the cases where males and females have different environmental optima. The error bar plots show the mean and standard deviation of 30 independent simulation realisations. }
\label{fig:FigureSIC1}
\end{figure}

In Figure 2 of the main text, we showed the evolved dispersal probability of males and females at equilibrium under static and homogeneous environment across habitat patches.
The results are similar under the other 3 types of environment, as shown in Figure \ref{fig:FigureSIC1}. 
The sex-specific equilibrium dispersal probability is featured with a male-bias under small $k_{LA}$ values throughout all cases. 

\subsubsection*{Effect of female choice on individual condition under different environment types}

\begin{figure}[H]
\begin{center}
\includegraphics[width=1\textwidth]{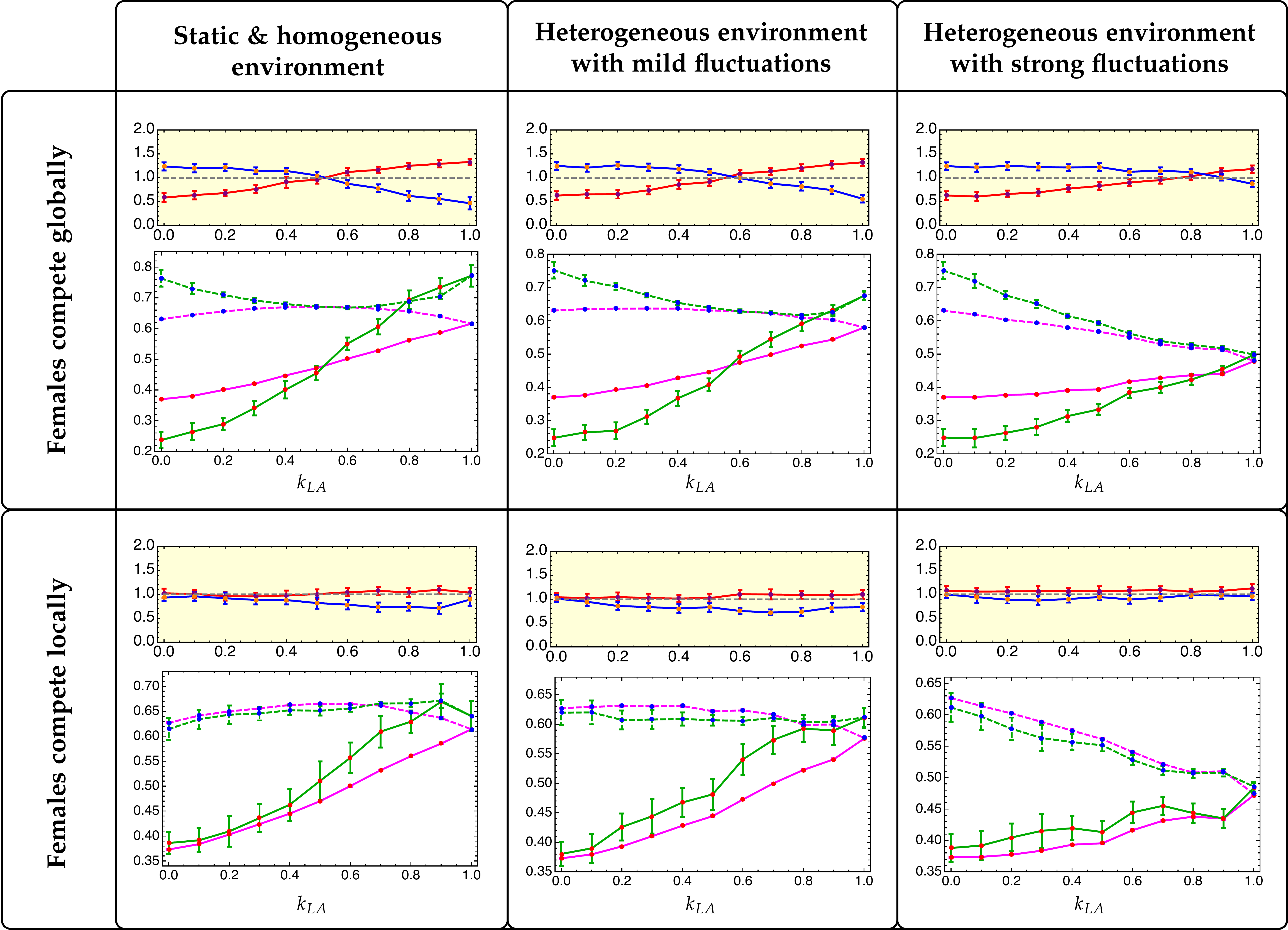}
\end{center}
\caption{Evolved $\alpha$ (red) and $\beta$ (blue) values and corresponding equilibrium condition of males (solid line and red marker) and females (dashed line and blue marker) when female choice is either allowed (green) or prevented intentionally (magenta), when females compete either globally or locally under 4 different types of environment. Each data point in the error plots and represents the mean and standard deviation of 30 independent simulations. Sexual conflict arises from IASC. }
\label{fig:FigureSIC2}
\end{figure}

In Figure 4 of the main text, we showed the effect of female choice on individual conditions when the environment is static but heterogeneous across habitat patches. 
Here we show in Figure \ref{fig:FigureSIC2} that qualitatively similar results also hold under different types of environment. 

As illustrated in Figure \ref{fig:FigureSIC2}, under global competition (hard selection), choosy females always have higher or equal condition than females that are forced to mate randomly. 
But when selection is soft, choosy females can be trapped in a ``tragedy of the commons'' scenario due to competition for producing attractive sons (via mating with high-condition males) when sexual conflict is strong ($k_{LA}$ is small).

\subsubsection*{Does it matter whether condition is determined in the natal or breeding habitat?}
\begin{figure}[H]
\begin{center}
\includegraphics[width=1\textwidth]{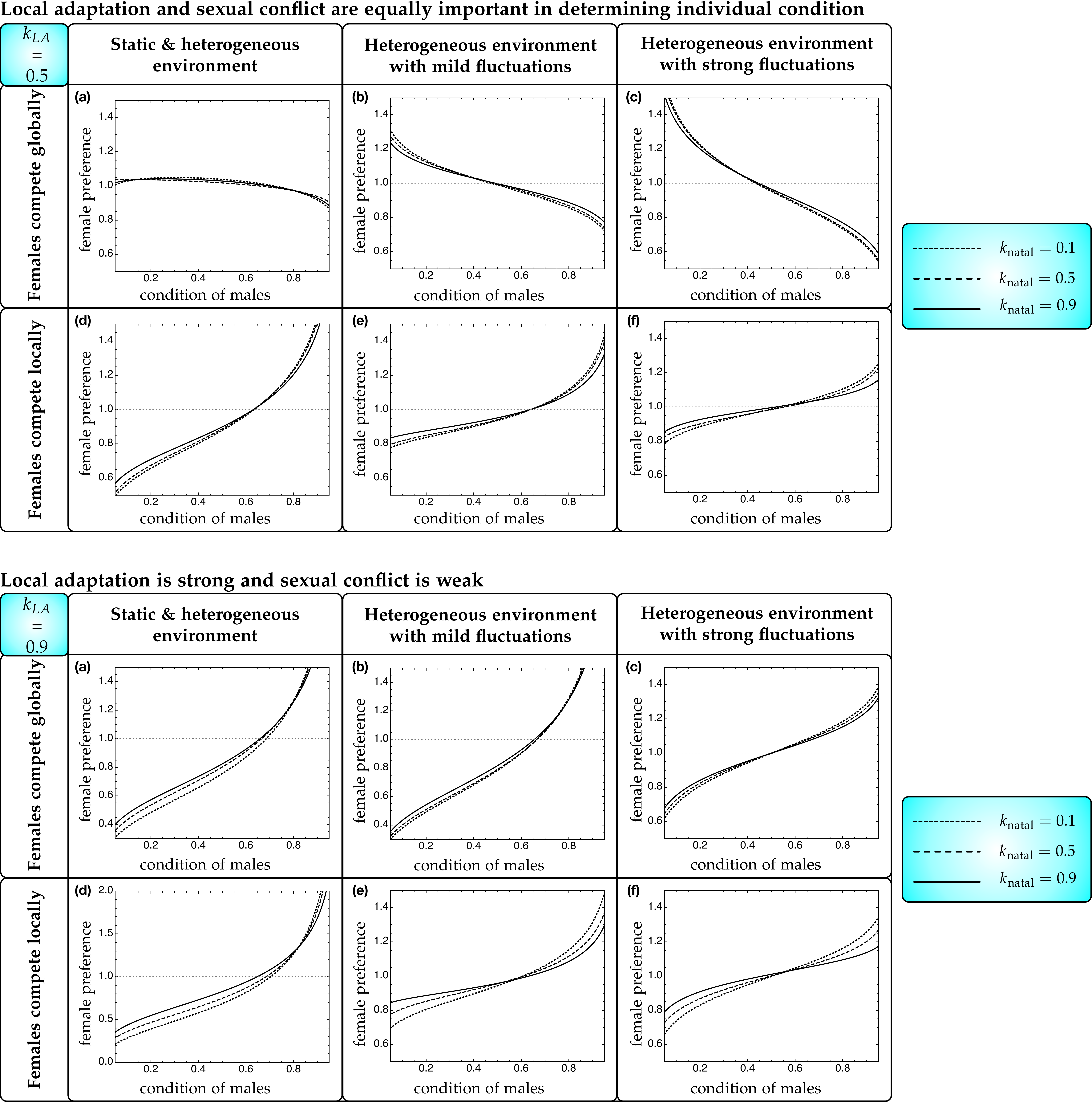}
\end{center}
\caption{The effect of the relative weight of the natal habitat in determining individual condition ($k_\text{natal}$) on female preference under different environments and different scale of female competition. Each curve is plotted using the mean $\alpha$ and  $\beta$ values calculated from the last 2000 generations of 30 independent realisations. In all simulations, sexual conflict arises from the IASC locus, and the sex-specific dispersal probabilities can coevolve with female choice. }
\label{fig:FigureSIC3}
\end{figure}

We find that slightly stronger female choice evolves when condition is determined primarily in the breeding patch rather than the natal patch (i.e. when $k_{\text{natal}}$ is low), irrespective of whether high-condition or low-condition males are preferred (Figure \ref{fig:FigureSIC3}).
This result is intuitive because with low  $k_{\text{natal}}$, females gain more information about adaptation to the environment of the breeding patch from male condition, which increases the benefit of condition-based choice, assuming that all else is equal and that most offspring remain in the breeding patch. 
As expected, the effect of $k_{\text{natal}}$ on female preference is larger when local adaptation, rather than IASC, is the main determinant of condition. 
As shown in Figure \ref{fig:FigureSIC3}, the impact of $k_{\text{natal}}$ is slightly larger when local adaptation plays a major role in determining individual condition ($k_{LA}=0.9$) than when local adaptation and sexual conflict are equally important ($k_{LA}=0.5$).

\end{document}